\documentclass[11pt,a4paper]{article}
\usepackage[utf8]{inputenc}
\usepackage[T1]{fontenc}
\usepackage{newtxtext,newtxmath}
\usepackage[margin=1in,top=1.1in,bottom=1.1in]{geometry}
\usepackage{microtype}
\usepackage{titlesec}
\usepackage{enumitem}
\usepackage{xcolor}
\usepackage{graphicx}
\usepackage{tikz}
\usetikzlibrary{shapes.geometric, arrows.meta, positioning, fit, calc, backgrounds, trees}
\usepackage[round,authoryear]{natbib}
\usepackage[colorlinks=true,linkcolor=red,citecolor=green,urlcolor=cyan]{hyperref}
\setcitestyle{open={[},close={]},sep={;},aysep={,}}

\selectfont
\titlespacing*{\section}{0pt}{1.2em}{0.5em}
\titlespacing*{\subsection}{0pt}{0.9em}{0.4em}
\setlist[itemize]{leftmargin=1.6em,itemsep=2pt,topsep=2pt}
\definecolor{skyblue}{HTML}{4AA3E3}

\begin{document}
\begin{center}
{\LARGE\bfseries AccountAgent: AI Accounting Assistant System}\\[0.6em]
{\normalfont\large Yulu Huang\footnotemark[2], Niannian Yu, Yaxin Yang, Jinpeng Lv}\\[0.3em]
{\normalfont\itshape Jiangxi University of Finance and Economics}\\[0.15em]
{\normalfont\footnotesize\textsuperscript{$\dagger$}Corresponding author: \texttt{yuluhuang324@gmail.com}}
\end{center}
\footnotetext[2]{Corresponding author.}
\vspace{0.4em}

\begin{quote}
\noindent\textbf{Abstract.} The AI Accounting Assistant System is an innovative tool that improves the accuracy and efficiency of financial management and is becoming a core support for enterprise accounting. It relies on machine learning, natural language processing, and data visualization to automate the full accounting agent including bookkeeping, report generation, and data analysis, substantially reducing manual operations and minimizing human error. Designed to resolve the pain points of low efficiency, cumbersome workflows, and data lag in traditional accounting, the system shifts financial work from repetitive labor toward high-value decision support. It deeply mines historical financial data, precisely identifies operating trends, and provides real-time, targeted insight for strategic planning, risk prevention, and operating decisions. By reconstructing the accounting agent, the system realizes automated bookkeeping, intelligent analysis, and efficient compliance, driving the accounting profession from a bookkeeping orientation toward a management orientation. This document presents the architecture, methodology, key algorithms, and functional modules of the platform.

\vspace{0.3em}
\noindent\textbf{Keywords:} AI Accounting, Voucher Processing, Business-Finance Integration, Anomaly Detection, Tax Compliance, Fund Forecasting, Multimodal Large Language Model
\end{quote}
\vspace{0.3em}
\section{Introduction}
Enterprise accounting is a foundational yet labor-intensive function whose accuracy and timeliness directly affect financial reporting and decision making. Traditional accounting workflows are constrained by low efficiency, cumbersome procedures, and data lag: vouchers are entered manually, business and financial data are disconnected, and analysis arrives too late to inform proactive decisions. As enterprises digitize, these limitations become a competitive bottleneck. \citep{yang2024,yang2006,lewis2020}

Artificial intelligence and large language models offer a path beyond manual bookkeeping. A multimodal large model can recognize vouchers across more than thirty receipt types, match accounting subjects using a domain knowledge base, and generate bookkeeping vouchers automatically, while integrated analysis engines support cash-flow forecasting, cost monitoring, and sales-trend analysis. When coupled with natural-language interaction and a cloud-native architecture, such a model shifts accounting from passive record-keeping toward active warning and decision support. \citep{brown2020,wu2023,yang2024}

The AI Accounting Assistant System realizes this vision through a comprehensive platform coupling a multimodal large model with business-finance integration, analysis, and compliance engines. Users interact through a conversational interface and the system executes voucher processing, cross-module data flow, multidimensional analysis, tax compliance, and fund forecasting in a single end-to-end pipeline. The platform targets enterprise finance departments and accounting firms. \citep{wu2023}

This document presents the overall design, the methodology, and the functional modules. The remainder is organized as follows: Section 2 reviews related work; Section 3 details the methodology; Section 4 presents the system functions; and Section 5 concludes. \citep{ouyang2022}

Accounting is the ledger of enterprise reality, and the integrity of that ledger underpins every downstream decision, from pricing and budgeting to investment and compliance. Yet the work of maintaining it, classifying vouchers, reconciling modules, and closing periods, remains disproportionately manual, slow, and error-prone. A platform that automates this work while preserving its accuracy does not merely improve an operational function; it elevates the reliability of the information foundation on which the enterprise is managed. \citep{touvron2023}

Constructing such a platform requires reconciling the flexibility of large-model reasoning with the precision that accounting demands. The system resolves this by coupling a multimodal voucher model with deterministic subject-matching and reconciliation engines, so that recognition, classification, and integration are fast yet auditable, and finance teams are freed to focus on analysis and strategy. \citep{dosovitskiy2021,wu2023,hochreiter1997}

Accounting is the ledger of enterprise reality, and the integrity of that ledger underpins every downstream decision, from pricing and budgeting to investment and compliance. Yet the work of maintaining it, classifying vouchers, reconciling modules, and closing periods, remains disproportionately manual, slow, and error-prone. A platform that automates this work while preserving its accuracy does not merely improve an operational function; it elevates the reliability of the information foundation on which the enterprise is managed, and a more reliable foundation yields better decisions throughout the organization. \citep{touvron2023}

Constructing such a platform requires reconciling the flexibility of large-model reasoning with the precision that accounting demands. The system resolves this by coupling a multimodal voucher model with deterministic subject-matching and reconciliation engines, so that recognition, classification, and integration are fast yet auditable, and finance teams are freed to focus on analysis and strategy rather than on the mechanical entry that consumes the majority of accounting labor. \citep{dosovitskiy2021,wu2023,hochreiter1997}

\section{Related Works}
Accounting automation has historically progressed through spreadsheet macros, optical-character-recognition pipelines, and rule-based bookkeeping engines. These systems reduced manual data entry but offered limited recognition accuracy on complex vouchers and limited support for cross-module business-finance integration, leaving substantial manual reconciliation. \citep{dosovitskiy2021,wu2023,lu2020}

Optical character recognition and document-ai research contributed high-accuracy recognition of invoices and receipts, while natural-language processing and large language models enabled subject matching and report generation. Vision-language models further strengthened multimodal voucher understanding. Financial-domain models such as FinBERT, FinGPT, and BloombergGPT demonstrated gains from domain adaptation, and retrieval-augmented generation grounded answers in up-to-date policy to reduce hallucination. \citep{dosovitskiy2021,brown2020,wu2023}

Despite these advances, integrating multimodal voucher processing, business-finance data integration, multidimensional analysis, tax compliance, and fund forecasting within a single accounting platform remains challenging. The system presented in this document addresses this gap by coupling a multimodal large model with integration and analysis engines behind a conversational interface, delivering end-to-end accounting intelligence. \citep{wu2023}

Business-process integration and continuous-close research form a further relevant tradition. Enterprise resource planning systems and robotic process automation have long sought to connect transactions to accounting automatically, yet the last mile of voucher generation and period close has resisted full automation. Multimodal recognition and large-model classification now close this gap, motivating the system's business-finance integration and voucher modules. \citep{dosovitskiy2021,wu2023,yang2024}

Business-process integration and continuous-close research form a further relevant tradition. Enterprise resource planning systems and robotic process automation have long sought to connect transactions to accounting automatically, yet the last mile of voucher generation and period close has resisted full automation. Multimodal recognition and large-model classification now close this gap, motivating the system's business-finance integration and voucher modules. \citep{dosovitskiy2021,wu2023,yang2024}

Anomaly detection and continuous-control monitoring constitute a parallel tradition. Statistical and machine-learning methods have long flagged unusual transactions for review, and continuous-control frameworks have advocated real-time monitoring over periodic audit. The system's analysis-and-warning module draws on these foundations, surfacing anomalies as they arise rather than at period end, when correction is costlier and the window for intervention has closed. \citep{hochreiter1997,liu2012,touvron2023}

\section{Methodology}
The system adopts a cloud-native and microservices four-layer architecture that realizes modularization and extensibility. The visual encoder processes image portions with a fourteen-by-fourteen patch size across twenty-four blocks, followed by a two-layer linear multilayer-perceptron projection that produces compressed visual tokens. Visual tokens and text tokens are finally passed through a decoder large language model for next-token prediction learning, realizing joint multimodal understanding. \citep{dosovitskiy2021,liu2023,brown2020}

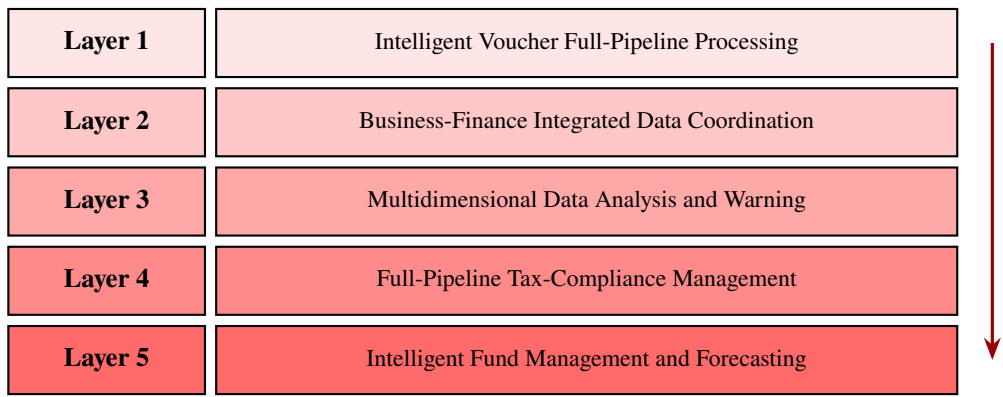
\begin{figure}[ht]
\centering
\begin{tikzpicture}[
  lbox/.style={draw, minimum width=2.6cm, minimum height=0.9cm, align=center, font=\small\bfseries, thick},
  cbox/.style={draw, minimum width=9.8cm, minimum height=0.9cm, align=center, font=\footnotesize, thick},
  barrow/.style={-Stealth, very thick, red!60!black}]

\node[lbox, fill=red!10] (la0) {Layer 1};
\node[cbox, fill=red!10, right=0.12cm of la0] (ca0) {Intelligent Voucher Full-Pipeline Processing};
\node[lbox, fill=red!22, below=0.12cm of la0] (la1) {Layer 2};
\node[cbox, fill=red!22, below=0.12cm of ca0] (ca1) {Business-Finance Integrated Data Coordination};
\node[lbox, fill=red!34, below=0.12cm of la1] (la2) {Layer 3};
\node[cbox, fill=red!34, below=0.12cm of ca1] (ca2) {Multidimensional Data Analysis and Warning};
\node[lbox, fill=red!46, below=0.12cm of la2] (la3) {Layer 4};
\node[cbox, fill=red!46, below=0.12cm of ca2] (ca3) {Full-Pipeline Tax-Compliance Management};
\node[lbox, fill=red!58, below=0.12cm of la3] (la4) {Layer 5};
\node[cbox, fill=red!58, below=0.12cm of ca3] (ca4) {Intelligent Fund Management and Forecasting};
\draw[barrow] ($(ca0.east)+(0.45,0)$) -- ($(ca4.east)+(0.45,0)$);
\end{tikzpicture}
\caption{Layered architecture of the AI Accounting Assistant System, shown as a two-column layer map. Each layer label maps to its constituent module; the side arrow indicates top-down control flow across layers.}
\label{fig:arch}
\end{figure}
The multimodal large model recognizes more than thirty receipt types including value-added-tax invoices, bank receipts, and expense-reimbursement forms with accuracy above ninety-nine percent. It automatically extracts amounts, tax rates, and suppliers, matches accounting subjects against an industry knowledge base, and generates bookkeeping vouchers, supporting batch upload and review with monthly error rates driven below three-tenths of a percent. \citep{lewis2020}

\begin{figure}[ht]
\centering
\resizebox{\textwidth}{!}{%
\begin{tikzpicture}[
  pbox/.style={draw, rounded corners=2pt, align=center, minimum width=2.0cm, minimum height=0.85cm, font=\footnotesize, thick, fill=blue!14},
  lbox/.style={draw, align=center, minimum width=2.4cm, minimum height=1.5cm, font=\footnotesize, thick, fill=blue!32, rounded corners=4pt},
  sbox/.style={draw, rounded corners=2pt, align=center, minimum width=2.0cm, minimum height=0.85cm, font=\footnotesize, thick, fill=orange!12},
  arrow/.style={-Stealth, thick, line width=0.9pt, blue!55!black}]

\node[pbox] (p0) {Receipt/Entry};
\node[pbox, right=0.3cm of p0] (p1) {Classify};
\draw[arrow] (p0) -- (p1);
\node[pbox, right=0.3cm of p1] (p2) {Reconcile};
\draw[arrow] (p1) -- (p2);
\node[lbox, right=0.3cm of p2] (p3) {LLM Verify};
\draw[arrow] (p2) -- (p3);
\node[sbox, right=0.3cm of p3] (p4) {Report};
\draw[arrow] (p3) -- (p4);
\node[pbox, right=0.3cm of p4] (p5) {File};
\draw[arrow] (p4) -- (p5);
\draw[arrow, dashed] (p5.north) to[bend right=40] (p0.north);
\end{tikzpicture}
}
\caption{Processing pipeline of the AI Accounting Assistant System. Inputs flow left-to-right through each stage; the dashed arc denotes a feedback loop back to the start.}
\label{fig:pipeline}
\end{figure}
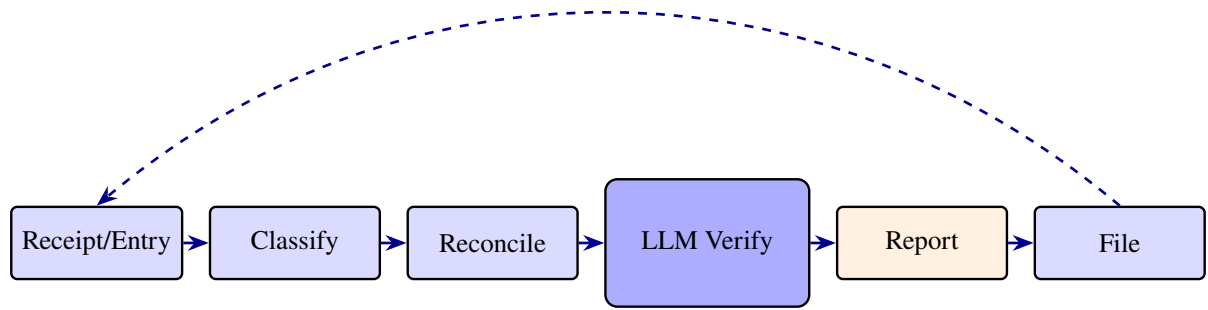
A policy dynamic-adaptation engine uses retrieval-augmented generation to monitor tax and accounting-standards updates in real time, automatically updating compliance-check rules and declaration templates so that the enterprise always complies with current requirements. The system is elastic and horizontally scalable, supporting growth from thousands to tens of thousands of vouchers per month, and supports on-demand module activation for enterprises at different stages. \citep{lewis2020,hu2022,chen2016}

The multimodal voucher model is tuned to the long tail of receipt variety. Beyond standard invoices, real enterprises process bank receipts, expense forms, customs declarations, and hand-written vouchers whose layouts differ widely. A fourteen-by-fourteen patch vision encoder captures fine print and seal detail, a two-layer projection compresses visual tokens, and a decoder language model trained on next-token prediction unifies visual and textual understanding, achieving recognition accuracy above ninety-nine percent across more than thirty receipt types. \citep{dosovitskiy2021,liu2023,brown2020}

A policy dynamic-adaptation engine is essential in a domain where rules change. Retrieval-augmented generation monitors tax-authority and accounting-standards publications in real time, and when a rate or rule changes, the system updates its compliance-check rules and declaration templates automatically and alerts affected workflows. This keeps the enterprise continuously compliant without the lag that produces risk, and an elastic, horizontally scalable architecture supports growth from thousands to tens of thousands of vouchers per month. \citep{lewis2020,hu2022,chen2016}

The multimodal voucher model is tuned to the long tail of receipt variety. Beyond standard invoices, real enterprises process bank receipts, expense forms, customs declarations, and hand-written vouchers whose layouts differ widely. A fourteen-by-fourteen patch vision encoder captures fine print and seal detail, a two-layer projection compresses visual tokens, and a decoder language model trained on next-token prediction unifies visual and textual understanding, achieving recognition accuracy above ninety-nine percent across more than thirty receipt types. \citep{dosovitskiy2021,liu2023,brown2020}

A policy dynamic-adaptation engine is essential in a domain where rules change. Retrieval-augmented generation monitors tax-authority and accounting-standards publications in real time, and when a rate or rule changes, the system updates its compliance-check rules and declaration templates automatically and alerts affected workflows. This keeps the enterprise continuously compliant without the lag that produces risk, and an elastic, horizontally scalable architecture supports growth from thousands to tens of thousands of vouchers per month. \citep{lewis2020,hu2022,chen2016}

Human-in-the-loop checkpoints preserve accountability without sacrificing speed. High-stakes operations such as the finalization of a period close or the submission of a tax return require human confirmation, and the system presents the supporting evidence and computed figures for review at each such checkpoint, ensuring that automation accelerates the accountant without removing the judgment that control and compliance require, and without creating an unreviewed path from data to filing. \citep{ouyang2022}

Recognition and classification accuracy are maintained through continuous evaluation against auditor-confirmed ground truth. Each voucher classification is paired with its eventual confirmation or correction, and per-receipt-type accuracy is tracked, so that drift in a long-tail receipt type is detected and corrected before it corrupts the ledger, preserving the integrity on which all downstream analysis depends. \citep{dosovitskiy2021,hochreiter1997,lu2020}

A retrieval layer over accounting standards and tax policy grounds the model's classification and compliance reasoning in citable authority rather than in parametric memory alone. When the model classifies a voucher or validates a declaration, the relevant standards and current policy are retrieved and injected into its context, so that classifications cite governing rules and compliance checks reflect the regulation in force for the filing period, preventing the silent non-compliance that a static, memorized rule set would produce as policy evolves. \citep{lewis2020,hochreiter1997,wei2022}

Serving is organized around the accounting task rather than around the model in isolation. A task graph parsed from the user's query orchestrates voucher recognition, business-finance integration, anomaly analysis, tax compliance, and fund forecasting as coordinated subtasks, with the model invoked where recognition and interpretation are required and deterministic engines invoked where computation is required, keeping each figure traceable to its source and preventing the model from fabricating the numerical results on which a defensible close or filing depends. \citep{dosovitskiy2021,wu2023,yang2024}

Security and access control are integral to a platform handling the enterprise ledger. Role-based access restricts voucher and financial data to authorized finance roles, every action is logged with actor and timestamp, and the separation between entry and approval is enforced by the access model, so that no single role can both record and approve a transaction, preserving the control principle that accounting and audit require even as automation accelerates the close. \citep{wu2023,yang2024,yang2006}

Observability is built into the pipeline. Every recognized voucher, classification, and reconciliation is traced, so that the provenance of any figure in a report or filing can be reconstructed, supporting both internal control review and the external audit scrutiny that financial reporting invites, and ensuring the platform's intelligence is as inspectable as the records it produces. \citep{yang2024,yang2006}

The platform's evaluation discipline extends to the qualitative as well as the quantitative. Beyond tracking recognition accuracy, the model's classifications and anomaly interpretations are assessed for grounding against auditor-confirmed ground truth, so that a classification is held to a standard as strict as the recognition that produced it, preventing the failure in which a correctly read figure is misclassified and propagates through the ledger as a silent error. \citep{dosovitskiy2021,liu2012,lu2020}

\section{System Function}
The platform provides five core functional modules covering intelligent voucher full-pipeline processing, business-finance integrated data coordination, multidimensional data analysis and warning, full-pipeline tax-compliance management, and intelligent fund management and forecasting. Each module is described below.

The five modules form a closed accounting loop rather than a linear chain. Recognized vouchers drive business-finance integration; integration surfaces the anomalies that analysis flags; the analysis conditions the tax-compliance checks; and the fund forecast's gaps redirect attention to the modules that can close them. This closure is what transforms the platform from a set of automation tools into a financial-control instrument that prevents issues rather than reporting them.

\subsection{Intelligent Voucher Full-Pipeline Processing}
Using character recognition and multimodal technology, the system realizes automated processing of multiple voucher types, accurately recognizing more than thirty receipt types including value-added-tax invoices, bank receipts, and expense-reimbursement forms with accuracy above ninety-nine percent. It automatically extracts amounts, tax rates, and suppliers, matches accounting subjects against an industry knowledge base, and generates bookkeeping vouchers, supporting batch upload and review with monthly error rates driven below three-tenths of a percent.

Voucher recognition is robust to the imperfections of real documents. Smudged ink, skewed scans, and mixed-language fields are handled by the multimodal model in concert with layout-aware preprocessing, so that recognition accuracy remains high across the degraded inputs that production accounting actually encounters, rather than only across the clean inputs that benchmark evaluation implies.

Voucher recognition is robust to the imperfections of real documents. Smudged ink, skewed scans, and mixed-language fields are handled by the multimodal model in concert with layout-aware preprocessing, so that recognition accuracy remains high across the degraded inputs that production accounting actually encounters, rather than only across the clean inputs that benchmark evaluation implies.

\begin{itemize}
  \item Multi-type recognition: thirty-plus receipt types with above-ninety-nine-percent accuracy
  \item Subject matching and voucher generation: knowledge-base-driven accounting-subject matching
  \item Batch processing: batch upload and review with sub-percent error rates
\end{itemize}
\subsection{Business-Finance Integrated Data Coordination}
Through intelligent algorithms the system breaks business and financial data barriers to realize real-time cross-module data flow. After a sales order is confirmed an accounts-receivable voucher is automatically generated, and after purchase warehousing an accounts-payable process is triggered, without manual data transfer. The system connects to inventory, reimbursement, and fund-management modules, compressing monthly closing time from days to a single day.

Business-finance integration is event-driven rather than batch-scheduled. The generation of an accounts-receivable voucher follows the confirmation of a sales order immediately, not at a nightly reconciliation, so that the ledger reflects the business in near real time and period close becomes a confirmation rather than a reconstruction, compressing the close cycle from days to hours.

Business-finance integration is event-driven rather than batch-scheduled. The generation of an accounts-receivable voucher follows the confirmation of a sales order immediately, not at a nightly reconciliation, so that the ledger reflects the business in near real time and period close becomes a confirmation rather than a reconstruction, compressing the close cycle from days to hours.

\begin{itemize}
  \item Automated voucher triggering: accounts-receivable and accounts-payable auto-generation
  \item Cross-module coordination: inventory, reimbursement, and fund-management integration
  \item Closing acceleration: monthly closing compressed from days to one day
\end{itemize}
\subsection{Multidimensional Data Analysis and Warning}
Leveraging large-model analysis, the system builds prediction models from historical data to realize cash-flow forecasting, cost-volatility monitoring, and sales-trend analysis. A visualized operating dashboard presents core indicators, automatically identifies data anomalies such as sudden product-line profit-margin declines, and pushes warnings, supporting drill-down from enterprise-wide profit to single-category profit.

Anomaly detection is contextual rather than absolute. A figure that is unremarkable in one product line or season may be anomalous in another, and the system interprets each indicator against its own baseline, so that warnings reflect genuine deviations from expected behavior rather than generic thresholds, and so that attention is drawn to the deviations that matter rather than to those that merely exceed a fixed limit.

Anomaly detection is contextual rather than absolute. A figure unremarkable in one product line or season may be anomalous in another, and the system interprets each indicator against its own baseline, so that warnings reflect genuine deviations from expected behavior rather than generic thresholds, and attention is drawn to the deviations that matter rather than to those that merely exceed a fixed limit.

\begin{itemize}
  \item Prediction models: cash-flow, cost-volatility, and sales-trend forecasting
  \item Visualized dashboard: core-indicator presentation and anomaly identification
  \item Drill-down analysis: from enterprise profit to single-category profit
\end{itemize}
\subsection{Full-Pipeline Tax-Compliance Management}
The system automatically aggregates output and input tax data, computes tax payable, and generates standardized declaration forms. After connecting to the electronic tax bureau system it supports one-click upload with simultaneous filing-receipt archiving. A policy-matching function validates data against the latest tax policy, flags adjustment items, and after confirmation one-click declaration completes filing with the electronic tax bureau.

Tax compliance is prospective rather than retrospective. The policy-matching engine validates data against the regulations that will govern the filing period, not merely those in force at the moment of validation, so that a rate change effective on the filing date is respected before submission rather than discovered after, preventing the avoidable non-compliance that a snapshot-based check would permit.

Tax compliance is prospective rather than retrospective. The policy-matching engine validates data against the regulations that will govern the filing period, not merely those in force at the moment of validation, so that a rate change effective on the filing date is respected before submission rather than discovered after, preventing the avoidable non-compliance that a snapshot-based check would permit.

\begin{itemize}
  \item Tax aggregation and computation: output and input data with payable computation
  \item Declaration and archiving: standardized forms, one-click upload, and receipt archiving
  \item Policy matching: compliance validation against latest tax policy
\end{itemize}
\subsection{Intelligent Fund Management and Forecasting}
The model possesses strong analytical and forecasting capability, learning the distribution of large volumes of historical data to forecast fund gaps for the next thirty days and output a health-score report, triggering automatic warning when the score falls below seventy. Application substantially improves fund-forecast accuracy and reduces annual fund cost.

Fund forecasting distinguishes structure from noise. The model decomposes a cash-flow series into trend, seasonality, and residual, and forecasts each component on its appropriate horizon, so that a predicted gap reflects a genuine structural shortfall rather than normal seasonality, and a health score communicates the severity of a gap in terms that drive action rather than anxiety.

Fund forecasting distinguishes structure from noise. The model decomposes a cash-flow series into trend, seasonality, and residual, and forecasts each component on its appropriate horizon, so that a predicted gap reflects a genuine structural shortfall rather than normal seasonality, and a health score communicates severity in terms that drive action rather than anxiety.

\begin{itemize}
  \item Gap forecasting: thirty-day fund-gap prediction from historical distributions
  \item Health scoring: fund-health-score reports with automatic low-score warning
  \item Cost reduction: improved forecast accuracy lowering annual fund cost
\end{itemize}
\section{Conclusion}
The AI Accounting Assistant System reconstructs the accounting agent through a multimodal large model, realizing automated bookkeeping, intelligent analysis, and efficient compliance that shift accounting from a bookkeeping orientation toward a management orientation. The cloud-native microservices architecture, the multimodal voucher model, and the policy dynamic-adaptation engine together form a robust foundation for enterprise financial intelligence.

The system's functional completeness, spanning voucher processing, business-finance integration, analysis and warning, tax compliance, and fund forecasting, substantially reduces manual operations and human error and improves financial efficiency and accuracy. A human-machine collaborative model lets finance teams focus on high-value work such as cost analysis and strategic planning.

Future work will extend the system to continuous real-time accounting, cross-entity consolidated reporting, and proactive financial-risk prevention, further evolving the platform into an indispensable intelligent brain for enterprise accounting and financial management.

As the platform evolves toward continuous real-time accounting and cross-entity consolidated reporting, the boundary between recording and managing will dissolve, positioning the system as the central nervous system of enterprise finance and the foundation of a genuinely proactive financial-control function.

As the platform evolves toward continuous real-time accounting and cross-entity consolidated reporting, the boundary between recording and managing will dissolve, positioning the system as the central nervous system of enterprise finance and the foundation of a genuinely proactive financial-control function that prevents issues rather than reporting them after the fact.

The platform's ultimate contribution is the elevation of the accounting function itself. By absorbing the mechanical dimensions of entry, reconciliation, and close, the system reallocates professional attention to the analysis and strategy that generate marginal value, and in doing so it advances the profession from a bookkeeping orientation toward a management orientation, a transition the discipline has long sought and that this platform makes operationally tractable.

In sum, the platform's contribution is the elevation of the accounting function, advancing the profession from a bookkeeping orientation toward a management orientation by absorbing the mechanical dimensions of entry, reconciliation, and close, and reallocating professional attention to the analysis and strategy that generate marginal value, a transition the discipline has long sought.

Ultimately, the platform reframes the accountant's relationship with the ledger. By absorbing the mechanical dimensions of entry, reconciliation, and close while preserving the integrity on which all analysis depends, it allows the accountant's attention to concentrate on the judgment that remains irreducibly human, the interpretation of what the figures mean and the strategy they imply, and it is this elevation of altitude, rather than the automation of entry alone, that constitutes the platform's value.

\renewcommand{\refname}{References}

\end{document}